\documentclass[         %
aps,                    
prd,                    
showpacs,               
superscriptaddress,    
nofootinbib,            
onecolumn,             %
showkeys,               %
preprintnumbers,        %
]{revtex4}               
\usepackage{graphicx,amsmath}
\usepackage[justification=RaggedRight]{caption}
\usepackage{listings}
\usepackage{enumerate}
\usepackage{natbib}
\usepackage{xcolor}
\usepackage{float}
\usepackage[section]{placeins}
\usepackage{subfigure}
\usepackage{accents}
\usepackage{multirow}
\usepackage[utf8]{inputenc}
\definecolor{commentColor}{rgb}{0.84,0.10,0.11}

\begin{document}
\lstset{language=Mathematica,
  backgroundcolor=\color{white},   
  basicstyle=\footnotesize\ttfamily, 
  breaklines=true,                 
  commentstyle=\color{commentColor},
  escapeinside={~*}{*~},          
  extendedchars=true,              
  frame=single,                    
  keepspaces=true,                 
  numbers=left,                    
  numbersep=5pt,                   
  rulecolor=\color{black},         
  stepnumber=2,                    
  tabsize=2, }  
\title{Uncovering the Matter-Neutrino Resonance}
\author{D.~Väänänen}
\author{G. C. McLaughlin}
\affiliation{Department of Physics, North Carolina State University, Raleigh, NC 27695 USA.}
\date{\today}
\begin{abstract}

Matter Neutrino Resonances (MNRs) can drastically modify neutrino flavor evolution in astrophysical environments and may significantly impact nucleosynthesis. Here we further investigate the underlying physics of MNR type flavor transitions. We provide generalized resonance conditions and make analytical predictions for the behavior of the system. We discuss the adiabatic evolution of these transitions, considering both Symmetric and Standard scenarios.  {\it Symmetric} MNR transitions differ from {\it Standard} MNR transitions in that both neutrinos and antineutrinos can completely transform to other flavors simultaneously. We provide an example of the simplest system in which such transitions can occur with a neutrino and an antineutrino having a single energy and emission angle. We further apply linearized stability analysis to predict the location of self-induced nutation type (or bipolar) oscillations due to $\nu\nu$ -- interactions in the regions where MNR is ineffective. In all cases, we compare our analytical predictions to numerical calculations.
\end{abstract}
\medskip
\maketitle

\section{Introduction}
\label{sec:intro}

Compact object mergers and core collapse supernovae release a significant fraction of their energy in the form of neutrinos, e.g. \cite{Palenzuela:2015dqa,Dessart:2008zd,Deaton:2013sla,Perego:2014fma,O'Connor:2014rva,Abdikamalov:2012zi,Tamborra:2014hga}.  These neutrinos play a number of roles, for example they are a key player in determining the types of nucleosynthesis that come from the winds in these objects, e.g. \cite{Surman:2005kf,Roberts:2010wh}.  Developing further understanding of neutrino flavor evolution in these environments is necessary for gaining a complete picture of the nature of neutrino evolution in astrophysical settings.  In addition, an understanding of the flavor content of the neutrino spectrum is essential for determining whether the suitable conditions for the synthesis of various types of heavy elements can be met, e.g. \cite{Duan:2010af,Malkus:2012ts} as well as for extracting the most information from future observations.  

Neutrino flavor evolution in dense astrophysical environments is inherently a complex quantum many-body problem~\cite{Sigl:1992fn,Balantekin:2006tg,Friedland:2006ke,Cardall:2007zw,Fidler:2011yq,Volpe:2013jgr, Vaananen:2013qja, Vlasenko:2013fja, Serreau:2014cfa}. Neutrinos are known to be able to transform from one type to another as they propagate freely in space.
Interactions with the particles in the surrounding environment can drastically modify their flavor evolution with significant consequences. The MSW resonance conversion effect is known to be responsible for the neutrino transitions in the Sun, solving the long standing solar neutrino problem~\cite{MSW, ResEnhance, SNO, kamland_msw}. Neutrino interactions with other surrounding neutrinos can become important in environments with extremely high neutrino densities, such as in supernovae or in accretion disks above merging compact objects. In these environments, neutrino-neutrino potential can induce collective nutation type flavor transformation effects distinct from the MSW effect~\cite{collective,Duan:2010bg,Duan:2010bf}. At the highest densities, the quantum many-body nature of the problem can break several commonly utilized assumptions leading to a possibility of novel effects~\cite{Volpe:2013jgr, Vaananen:2013qja,Cherry:2013mv, Serreau:2014cfa, Vlasenko:2014bva, Cirigliano:2014aoa, Kartavtsev:2015eva}. 

In this manuscript, we concentrate on a new type of neutrino flavor transformation effect which was observed in the recent simulations of neutrino flavor evolution above black hole accretion disks~\cite{Malkus:2012ts}. 
Accretion disks produce mostly electron type neutrinos and antineutrinos. Antineutrinos have smaller emission disks with respect to neutrinos but are emitted with a hotter spectrum than that of neutrinos. Therefore, close to the neutrino emission disk antineutrino flux can dominate over neutrino flux. This is a distinct feature compared to standard proto-neutron star supernova neutrino scenarios. 

In Ref.~\cite{Malkus:2012ts}, two new types of neutrino transformations were observed which differ from the known MSW resonance and the self-induced bipolar transitions. In one of these transformations neutrinos fully convert to other flavors while antineutrinos return to their original configuration. This transformation was further investigated in Ref.~\cite{Malkus:2014iqa} and was understood to be a consequence of a {\it Matter Neutrino Resonance} (MNR) achieved by an active cancellation of the neutrino-neutrino potential and the background matter contribution. Hereafter, we refer to this type of transformation as {\it Standard} MNR transition. Another type of transformation that was observed in~\cite{Malkus:2012ts} fully converted both neutrinos {\it and} antineutrinos symmetrically. In Ref.~\cite{Malkus:2015mda} numerical studies were performed examining both Standard and Symmetric MNR and exploring their consequences for nucleosynthesis. Similar effects can occur in other non-linear systems that exhibit similar features, such as in the presence of neutrino-antineutrino spin coherence~\cite{Vlasenko:2014bva} or active-sterile neutrino mixing.

In order to fulfill the MNR condition, the neutrino-neutrino and the background matter potentials are required to have opposite signs. A characteristic feature of MNR is that the system can maintain the resonance over extended period by suitably transforming neutrinos to other flavors. 
In this manuscript, we present 
analytical conditions for the occurrence of MNR by deriving general resonance and adiabaticity conditions. We further explain the underlying physical mechanism as an adiabatic evolution of (anti)neutrino in-medium energy eigenstates. The analytical expressions describing (anti)neutrino survival probabilities during MNR transition are obtained by minimizing the 
equation governing the separation of the in-medium energy eigenstates. In the presence of large background matter, when MNR transition occurs, the analytical expressions for the survival probabilities have only a small dependence on vacuum mixing parameters. %
However, as discussed in \cite{Malkus:2014iqa} the presence of MNR transitions does depend on the values of these parameters.  Small vacuum mixing angles of steeply changing potentials will prevent MNR transitions from occurring. 

In the absence of (or in addition to) MNR transitions, neutrinos can undergo collective self-induced nutation type (or {\it  bipolar}) neutrino flavor transitions. Analytical conditions for the occurrence of self-induced effects have been obtained for two-neutrino systems with simplified (single-angle) geometrical dependence~\cite{Hannestad:2006nj,Raffelt:2007xt,Dasgupta:2007ws,Duan:2010bf, Galais:2011jh,Pehlivan:2011hp}. The onset of these self-induced effects has been related to the presence of an instability. In the context of supernova neutrinos the idea of employing linearized equations was first pointed out in Ref.~\cite{Sawyer:2008zs}.
This idea was further developed in Ref.~\cite{Banerjee:2011fj} where stability conditions were derived for the self-induced nutation type flavor transitions  
allowing to study conditions for the multi-emission angle effects in the case of two neutrino flavors. 
Thereafter, linearized stability analysis has been employed in several works, studying the suppression of collective effects during the accretion phase~\cite{Sarikas:2011am,Saviano:2012yh}, the effects of realistic emission angular distributions~\cite{Mirizzi:2011tu,Mirizzi:2012wp,Sawyer:2015dsa}, the presence of spurious instabilities due to the numerical inputs~\cite{Sarikas:2012ad}, the effects of neutrino scattering outside the neutrinosphere~\cite{Sarikas:2012vb}, instabilities triggered by flavor oscillation modes~\cite{Duan:2013kba}, the effects of breaking the axial symmetry~\cite{Raffelt:2013rqa} and deleptonization asymmetry~\cite{Chakraborty:2014lsa}, temporal instabilities~\cite{Dasgupta:2015iia} and effects of small scale features~\cite{Chakraborty:2015tfa}. General linearized equations, applicable to arbitrary number of flavors and a general form of the Hamiltonian, were derived in Ref.~\cite{Vaananen:2013qja}. In this manuscript we will apply the linearization procedure described in Ref.~\cite{Vaananen:2013qja} to systems with MNR resonances. 

The manuscript is structured as follows. In Sec. \ref{sec:setup} we describe the two flavor monoenergetic model as well as the density matrix formalism that we will use.  In Sec. \ref{sec:results} we derive the resonance conditions and the analytical expressions for the survival probabilities, illustrating the predictive power of our results with systems that exhibit either Symmetric MNR transitions or Standard MNR transitions.  We then consider systems that have MNR resonances but not MNR transitions due to suppression associated with small mixing angles.  In these cases, we demonstrate the usefulness of utilizing linear stability analysis to predict traditional bipolar transition regions.

\section{The Set Up}
\label{sec:setup}

\subsection{The Model}

In the following, we will consider a system that can be described by two (anti)neutrino flavors, which are produced and emitted with a single energy and a single emission angle.  
Neutrinos are assumed to be produced with a specific flavor described by an interaction eigenstate (or {\it flavor} state) $| \nu_f \rangle \ (f = e, \mu$ or $\tau)$\footnote{For a discussion on applicability of this assumption see e.g. Refs.~\cite{Giunti:2006fr} and references therein.}. 
The propagation eigenstates in vacuum, $| \nu_{i,j} \rangle  \  (i,j = 1, 2)$, can be written in terms of the flavor eigenstates as
\begin{equation}
\label{eq:2f}
	\begin{aligned}
  | \nu_1 \rangle &= \cos\theta_V \, | \nu_e \rangle + \sin\theta_V \, | \nu_x \rangle \ , \\
	| \nu_2 \rangle &= -\sin\theta_V \, | \nu_e \rangle + \cos\theta_V \, | \nu_x \rangle \ ,
	\end{aligned}
\end{equation}
with the vacuum mixing angle, $\theta_V$, determining the relative proportionality of the states. We define the mass-squared splitting between the neutrino propagation eigenstates in vacuum as $\delta m^2 \equiv m^2_2 - m^2_1$. Similar equations hold for antineutrinos.

The evolution of the considered system can be described by solving the following equations of motion for neutrino, $\rho$, and antineutrino, $\bar{\rho}$, density matrices, respectively,\footnote{For a derivation and listing of underlying assumptions of the utilized approach see Ref.~\cite{Volpe:2013jgr}.}:
\begin{equation}
\label{eq:evoeqs}	
	{\rm i} \frac{{\rm d} {\rho}}{{\rm d} r} = \left[H, {\rho}\ \right] \ , \quad
	{\rm i} \frac{{\rm d} \bar{\rho}}{{\rm d} r} = \left[\bar{H}, \bar{\rho}\ \right] . 
\end{equation}
In flavor basis the neutrino and antineutrino density matrices are defined, respectively, as
\begin{equation}
\label{eq:rhos}
\begin{aligned}
	\rho =&
		\begin{pmatrix}
		\rho_{ee} & \rho_{ex} \\
		\rho_{xe} & \rho_{xx} 
	\end{pmatrix}
  =
	\begin{pmatrix}
		\left|a_{\nu_e}\right|^2 & a_{\nu_e}a_{\nu_x}^* \\
		a_{\nu_e}^*a_{\nu_x}& \left|a_{\nu_x}\right|^2 
	\end{pmatrix}
		\ , \\ 
		\bar{\rho} =&
		\begin{pmatrix}
			\bar{\rho}_{ee} & \bar{\rho}_{ex} \\
			\bar{\rho}_{xe} & \bar{\rho}_{xx} 
	\end{pmatrix}
  =
	\begin{pmatrix}
		\left|a_{\bar{\nu}_e}\right|^2 & a_{\bar{\nu}_e}a_{\bar{\nu}_x}^* \\
		 a_{\bar{\nu}_e}^*a_{\bar{\nu}_x} & \left|a_{\bar{\nu}_x}\right|^2 
	\end{pmatrix}
		\ ,
\end{aligned}
\end{equation}
where $a_{\ \accentset{(-)}{\nu}_f}$ is a probability amplitude of a (anti)neutrino being in a given configuration $f$.
 
The total neutrino Hamiltonian of the investigated system in the flavor basis, $H_F$, consists of the vacuum, $H_V$, the background matter, $H_e$, and the neutrino-neutrino interaction, $H_{\nu\nu}$, contributions: 
\begin{equation}
\label{eq:HF1}
	H_F = H_V + H_e + H_{\nu\nu} \ .
\end{equation}
The vacuum Hamiltonian is given by 
\begin{equation}
\label{eq:HV}
H_V=\frac{\Delta_V}{2}\left(
		\begin{array}{cc}
		-\cos2\theta_V & \sin2\theta_V\\
		\sin2\theta_V & \cos2\theta_V
		\end{array}
	\right) \ ,
\end{equation}
where $\Delta_V \equiv \delta m^2/(2 E)$ with neutrino mass-squared splitting in vacuum, $\delta m^2$, and neutrino energy, $E$. The background matter contribution can be written as  
\begin{equation}
H_e =\left(
		\begin{array}{cc}
			V_e & 0 \\
			0 & 0
		\end{array}
	\right),
\end{equation}
where $V_e$ is the electron potential arising from the difference between net electron and positron number densities.  
The neutrino-neutrino interactions that couple the evolution of the neutrino and antineutrino densities are described by
\begin{equation}
	H_{\nu\nu} = \mu_{\nu} \left( \rho - \alpha \bar{\rho}^*\, \right) \ \ ,
\end{equation}
with interaction strength, $\mu_{\nu}$, asymmetry factor, $\alpha$, defining the relative difference between the initial $\nu_e$ and $\bar{\nu}_e$ number fluxes and * indicating complex conjugation operation. For antineutrinos the total interaction Hamiltonian is given by
\begin{equation}
	\bar{H}_F = H_V - H_e - H_{\nu\nu}^* \ . 
\end{equation}
For a derivation of these contributions in the utilized formalism see Ref.~\cite{Volpe:2013jgr}. 

A diagonal contribution can always be extracted from the Hamiltonian without impacting the flavor evolution described by Eqs.~\eqref{eq:evoeqs}. By subtracting a common factor $1/2 \left(V_e + \mu (\rho_{ee} + \rho_{xx} - \alpha (\bar{\rho}_{ee} + \bar{\rho}_{xx})) \right) {\rm  Diag(1,1)}$ from Eq.~\eqref{eq:HF1}, the total flavor basis neutrino Hamiltonian can be written in a symmetrized form as
\begin{equation}
\label{eq:HF}
H_F = \frac{1}{2}\left(
		\begin{array}{cc}
		-\Delta_V \cos2\theta_V + V_e + V_{\nu} & \Delta_V\sin2\theta_V + V_{\nu}^{ex}\\
		\Delta_V\sin2\theta_V + V_{\nu}^{xe} & \Delta_V\cos2\theta_V - (V_e + V_{\nu})
		\end{array}
	\right),
\end{equation}
where 
\begin{equation}
\label{eq:Vnu}
	\begin{aligned}
		V_{\nu} &\equiv \mu_{\nu} \left(\rho_{ee} - \rho_{xx} - \alpha (\bar{\rho}_{ee} - \bar{\rho}_{xx} ) \right) \ , \\
		V_{\nu}^{ex} &\equiv 2\mu_{\nu} (\rho_{ex} - \alpha \bar{\rho}_{xe} ) \ .
	\end{aligned}
\end{equation}
The antineutrino Hamiltonian is obtained from the above expression by replacing $V_e \rightarrow - V_e$, $V_{\nu} \rightarrow - V_{\nu}$ and $V_{\nu}^{ex}  \rightarrow - V_{\nu}^{xe}$.

\section{Results}

\label{sec:results}

\subsection{Resonance and Adiabaticity Conditions}

First we will derive equations to predict the location of the {\it Matter-Neutrino Resonance} (MNR). Then we proceed to derive equations to predict the anticipated behavior of the flavor evolution at the resonances by introducing a generalized {\it adiabaticity} parameter. 
The symmetrized flavor basis Hamiltonian, Eq.~\eqref{eq:HF}, can be diagonalized using a unitary $SU(2)$ rotation matrix $U_M$: 
\begin{equation}
\label{eq:HM}
	U_M^{\dagger} H_F U_M \equiv {\Delta_M} { \rm Diag} (-1,1) \equiv H_M \ ,
\end{equation}
 where $H_M$ is the instantaneous {\it in-medium} eigenbasis Hamiltonian with (anti)neutrino energy eigenstates $\pm\accentset{(-)\,}{\Delta}_M$. The eigen-energies define the ({\it effective}) in-medium mass-squared splitting, $\delta m^2_{eff}$:
\begin{equation}
\label{eq:deltaM}
\accentset{(-)}{\Delta}_M \equiv \frac{\delta m^2_{eff}}{2E} = \sqrt{\left(\Delta_V \cos2\theta_V\ \accentset{(+)}{-}\ (V_e + V_{\nu}) \right)^2 + 
	\left(\Delta_V \sin2 \theta_V\ \accentset{(-)}{+}\ V_{\nu}^{ex}\right)\left(\Delta_V \sin2 \theta_V\ \accentset{(-)}{+}\  V_{\nu}^{xe}\right) } \quad \ ,
\end{equation}
with parenthesis indicating the differences in case of antineutrinos.

The most general form of the rotation matrix (or {\it in-medium mixing matrix}) $U_M$ can be written as 
\begin{equation}
U_M = \left(
		\begin{array}{cc}
		1 & 0 \\
		0 & {\rm e}^{-{\rm i} \delta_M}
		\end{array}
	\right)
  \left(
		\begin{array}{cc}
		\cos\theta_M & \sin\theta_M\\
		-\sin\theta_M & \cos\theta_M
		\end{array}
	\right)
	\left(
		\begin{array}{cc}
		{\rm e}^{{\rm i} \beta_{1M}} & 0 \\
		0 & {\rm e}^{{\rm i} \beta_{2M}}
		\end{array}
	\right)
	\ .
\end{equation}
where $\theta_M$ is the in-medium mixing angle and $\delta_M, \beta_{1M,2M}$ are in-medium phases. The flavor composition of in-medium eigenstates, $| \nu_{iM} \rangle\ (i = 1,2)$, is obtained from the corresponding expression in vacuum, Eq.~\eqref{eq:2f}, by replacing the vacuum mixing angle, $\theta_V$, with the in-medium angle $\theta_M$.
The flavor Hamiltonian $H_F$ can then be written in terms of the in-medium quantities as 
\begin{equation}
\label{eq:HIM}
 H_F = U_M H_M U_M^{\dagger} = \frac{\Delta_M}{2} \left(
		\begin{array}{cc}
		-\cos2\theta_M & \sin2\theta_M {\rm e}^{{\rm i} \delta_M}\\
		\sin2\theta_M {\rm e}^{-{\rm i} \delta_M} & \cos2\theta_M
		\end{array}
	\right) \ .
\end{equation}

Notice that this expression is independent of the $\beta$ phases. The expressions for the flavor basis Hamiltonian in Eqs.~\eqref{eq:HF} and~\eqref{eq:HIM} give the following relations for the in-medium quantities:
\begin{equation}
\begin{aligned}
\Delta_M\cos{2\theta}_M &= \Delta_V \cos2\theta_V - (V_e + V_{\nu}) \ , \\
\Delta_M\sin{2\theta}_M {\rm e}^{{\rm i} \delta_M} &= \Delta_V \sin2\theta_V + V_{\nu}^{ex} \ .
\end{aligned}
\end{equation}
Combining the above relations one obtains the following equations for the in-medium mixing angle, $\theta_M$, and phase, $\delta_M$:
\begin{equation}
\label{eq:theta2M}
\begin{aligned}
\tan{2\theta}_M {\rm e}^{{\rm i} \delta_M} &= \frac{\tan2\theta_V + {\displaystyle \frac{ V_{\nu}^{ex}}{\Delta_V \cos2\theta_V} }}{1 - {\displaystyle \frac{V_e + V_{\nu}}{\Delta_V \cos2\theta_V}}} \ ,  \\
\tan\delta_M &= {\rm i}\ \frac{V_{\nu}^{ex} - V_{\nu}^{xe}}{2 \Delta_V \sin2\theta_V + V_{\nu}^{ex} + V_{\nu}^{xe}} = - \frac{{\cal I}m [V_{\nu}^{ex}]}{2 \Delta_V \sin2\theta_V + {\cal R}e [V_{\nu}^{ex}]} \ .
\end{aligned}
\end{equation}
The resonance condition is readily determined from above as:
\begin{equation}
\label{eq:MNRcond}
 (V_e + V_{\nu})|_{r=r_R} = \Delta_V \cos2\theta_V \ , 
\end{equation}
where $R$ indicates that the quantities are evaluated at the resonance location $r = r_R$. This equation allows one to determine the expected location of MNR.

Neutrino flavor evolution at a resonance depends on how the resonance is crossed. We quantify the crossing behavior by defining an {\it adiabaticity parameter}, $\gamma$.  The full expression for this parameter, including phase derivatives is given in Ref.~\cite{Galais:2011jh}.  In cases where the phase derivatives are unimportant, the adiabaticity parameter is given by:
\begin{equation}
\label{eq:gamma}
 \gamma \approx \frac{|\Delta_M|}{\left|\frac{\displaystyle{\rm d} \theta_M}{\displaystyle{\rm d r}}\right|} \ ,
\end{equation}
with the in-medium mass-squared splitting, $\Delta_M$ from Eq.~\eqref{eq:deltaM}, and rate of change of the in-medium mixing, $\left|{\displaystyle{\rm d} \theta_M}/{\rm d r}\right|$. If the rate of change of the in-medium mixing is much smaller with respect to the splitting of the energy eigenstates ($\gamma \gg 1$), neutrino stays on its in-medium eigenstate. In this case, we refer the evolution as being completely adiabatic. At the other extreme, if neutrino jumps to the other in-medium eigenstate, the evolution is said to be completely non-adiabatic ($\gamma \lesssim 1$). The adiabaticity of the resonance crossing in our 
systems can be evaluated by applying expressions for the in-medium quantities in Eqs.~\eqref{eq:deltaM} and~\eqref{eq:theta2M} to the definition of the full adiabaticity parameter from  Ref.~\cite{Galais:2011jh}.  At a location of the resonance the adiabaticity parameter becomes:
\begin{equation}
\label{eq:gammaR}
 \gamma_R = \left. \frac{2 \Delta_M^2 \sin 2\theta_{M}}{\displaystyle{\left|\frac{{\rm d} V_e}{\rm d r} + \frac{{\rm d}V_{\nu}}{\rm d r}\right|}} \right|_{r=r_R} \ . 
\end{equation}

\subsection{Analytical Expressions for the Survival Probabilities}

If a neutrino stays on its in-medium eigenstate one can write a probability of finding a neutrino with a flavor $f$ from the in-medium state, $| \nu_{iM} \rangle$, as
\begin{equation}
P(\nu_f | \nu_{iM}) \equiv |\langle \nu_f | \nu_{iM} \rangle|^2 \ .
\end{equation}
Therefore, the probability of finding e.g. an electron neutrino from in-medium state $ | \nu_{1M} \rangle$ can be expressed as
\begin{equation}
\label{eq:adiab}
\begin{aligned}
P(\nu_e | \nu_{1M}) = |\langle \nu_e | \nu_{1M} \rangle|^2 =& \cos^2{\theta}_M \\
=& \frac{1}{2} \left( 1 + \frac{\Delta_V \cos2 \theta_V - (V_e + V_{\nu})}{\Delta_M} \right) \\ 
=& \frac{1}{2} \left( 1 + \frac{1}{\displaystyle \sqrt{1 +  \left| \tan 2\theta_M \right|^2 } }  \right) \ ,
\end{aligned}
\end{equation}
where 
\begin{equation}
\label{eq:tan2tM}
  \left| \tan 2\theta_M \right|^2 = \frac{(\Delta_V \sin2 \theta_V + V_{\nu}^{ex})(\Delta_V \sin2 \theta_V + V_{\nu}^{xe})}{ \left( \Delta_V \cos2\theta_V - (V_e + V_{\nu}) \right)^2} \ .
\end{equation}

The separation of the in-medium neutrino energy eigenstates defines the in-medium mass splitting ``gap'' as
\begin{equation}
\label{eq:gap}
	G_M \equiv \Delta_M - (-\Delta_M) = 2\Delta_M \ ,
\end{equation}
where $\Delta_M$ is given by Eq.~\eqref{eq:deltaM}. Assuming that at some point during the evolution the in-medium eigenstates become very close, this ``gap'' reaches its minimum when $\Delta_M \approx 0$. By inspection of Eq. \ref{eq:deltaM}  one obtains two conditions, one for the flavor diagonal contributions:
\begin{equation}
\label{eq:diagmini}
	\Delta_V \cos2\theta_V - (V_e + V_{\nu}) \approx 0 \ ,
\end{equation}
and another one for the flavor off-diagonal contributions:
\begin{equation}
\label{eq:offdiagmini}
	\Delta_V \sin2 \theta_V + V_{\nu}^{ex} \approx 0 \quad ({\rm or}\quad \Delta_V \sin2 \theta_V + V_{\nu}^{xe} \approx 0) \ .
\end{equation}
The first condition, Eq.~\eqref{eq:diagmini}, is equivalent with the MNR condition Eq.~\eqref{eq:MNRcond}. The second condition, Eq.~\eqref{eq:offdiagmini}, generalizes the the observed MNR conditions in Ref.~\cite{Malkus:2014iqa} for the off-diagonal terms to include the vacuum contributions explicitly\footnote{In Ref.~\cite{Malkus:2014iqa} the conditions were written in  neutrino flavor isospin formalism~\cite{Duan:2005cp}: $s_{x,y} \approx -\alpha \bar{s}_{x,y}$}. Neglecting the vacuum corrections, in Ref.~\cite{Malkus:2014iqa} the above conditions were used to obtain analytical expressions for the electron (anti)neutrino survival probabilities when the MNR conditions are fulfilled:
\begin{equation}
\label{eq:anal}
\begin{aligned}
P_{\nu_e} \equiv \rho_{ee} &=  \frac{1}{2} \left(1 + \frac{\alpha^2 - R^2 - 1}{2 R} \right)  \ , \\
P_{\bar{\nu}_e} \equiv \bar{\rho}_{ee} &= \frac{1}{2} \left(1 + \frac{\alpha^2 + R^2 - 1}{2 \alpha R} \right) \ ,
\end{aligned}
\end{equation}
where $R \equiv V_e/\mu_{\nu}$ is the ratio of the neutrino-electron and neutrino-neutrino interaction scales. These equations were found to be in excellent agreement with the numerical results in Ref.~\cite{Malkus:2014iqa} during the MNR transition.

\begin{figure}[t!]
		\includegraphics[width=0.9\textwidth]{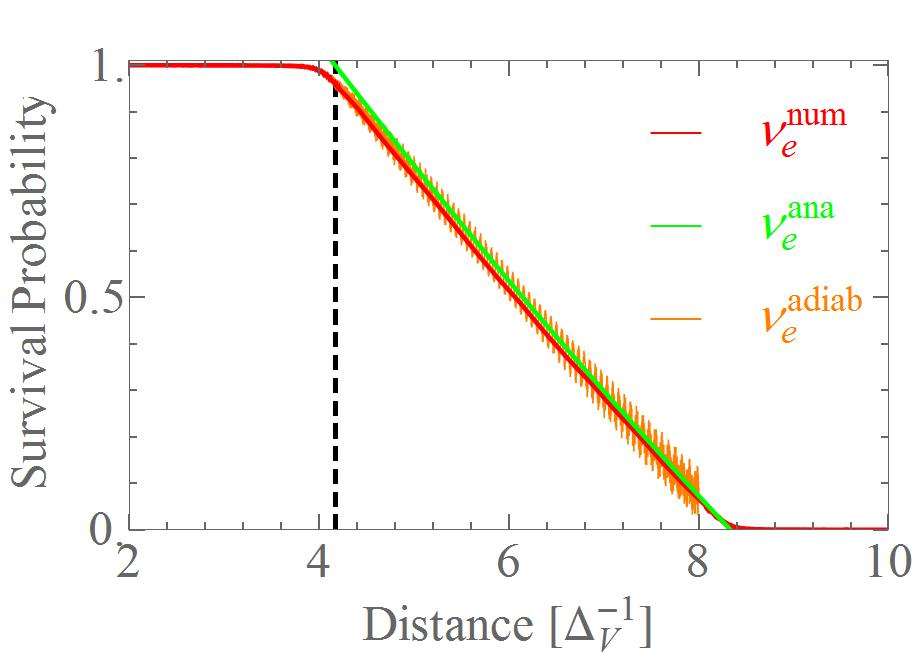}
  \caption{
	(Color online) Comparison of numerical results (red) for the $\nu_e$ survival probability with analytical prediction given by Eq.~\eqref{eq:anal} (green) and solution utilizing adiabatic assumption given by Eq.~\eqref{eq:adiab} (orange lines). The vertical dashed line represents the resonance location according to Eq.~\eqref{eq:MNRcond}. Here we have used the inverted neutrino mass hierarchy, $\Delta_V = -1$ (similar results hold for the normal mass hierarchy) and the mixing angle $\theta_V = 0.154$.}

	\label{fig:adiab}
\end{figure}

Fig.~\ref{fig:adiab} illustrates these results by comparing a numerically calculated Symmetric MNR transition with the adiabatic approximation, Eq.~\eqref{eq:adiab}, and the analytic prediction, Eq.~\eqref{eq:anal}.  In this figure we assume an inverted neutrino mass hierarchy, $\Delta_V = -1$, and a vacuum mixing corresponding to the measured value of $\theta_{13}$; $\theta_V = 0.154$.  We configure the system to capture the primary features of the symmetric MNRs seen in \cite{Malkus:2015mda}.  The neutrino potential starts negative and becomes positive, with an explicit parameterization of $\mu_{\nu} = 10\,000\ [\Delta_V]$ and the initial $\nu_e$ and $\bar{\nu}_e$ asymmetry factor, $\alpha(r) = a + b r$, with $a = 1.3$ and $b = -0.048$. The matter potential is kept constant at $V_e = 1000\ [\Delta_V]$. 

We see from this figure that the numerical results closely track adiabatic solution during most of the transition, demonstrating that the system remains in an instantaneous eigenstate throughout the bulk of the transition.  We see also that the analytic prediction stemming from the minimization of the ``gap'' between the instantaneous eigenstates also closely tracks the numerics.  We investigate this further in the next subsection and demonstrate that these results apply also to the Standard MNR transitions.

\subsection{Discussion of Symmetric and Standard MNR Transitions}
\label{sec:comp}

In this section we investigate in detail the results from the previous section in the context of two scenarios: one that produces a Standard MNR transition as described in Ref.~\cite{Malkus:2014iqa}, together with the simplest system exhibiting Symmetric MNR motivated by the results presented in Ref.~\cite{Malkus:2012ts} (see Region (III) in Figure 5 of~\cite{Malkus:2012ts}).  A characteristic feature of the Symmetric MNR region is that the initial neutrino-neutrino potential starts with a negative sign (antineutrinos dominate close to the neutrino emission) and changes its sign to positive (neutrinos dominate further out)~\cite{Malkus:2015mda}. This is different than the Standard MNR transition region where the neutrino-neutrino potential always remains negative~\cite{Malkus:2014iqa}. 

Therefore, we discuss two models in this section: A and B. The parameters of model A are chosen to capture the essential features of a Symmetric MNR transition and were also used in in Fig. \ref {fig:adiab}. Repeated here, they are $\mu_{\nu} = 10\,000\ |\Delta_V|$ and $\alpha(r) = a + b r$, with $a = 1.3$ and $b = -0.048$. The matter potential is kept constant at $V_e = 1000\ |\Delta_V|$. The parameters of model B are chosen to capture the essential features of a Standard MNR transition.  We use the same parameterization as in \cite{Malkus:2014iqa} for model B: $\mu_{\nu}(r) = 10\,000\ {\rm e}^{-r/10} |\Delta_V|$, $\alpha = 4/3$ and  $V_e = 1000\ |\Delta_V|$. In both of these models, $r$ represents distance in units of inverse $|\Delta_V|$. We have defined $\Delta_V$ so that it is $+1$ for the normal hierarchy and $-1$ for the inverted hierarchy. Consistent with available estimates for the placement of MNR in compact object mergers and core collapse supernova accretion disks \cite{Malkus:2012ts,Malkus:2015mda}, we have chosen models where $\mu_\nu >> \Delta_V$ throughout the transition.  

\begin{table}[h]
	\begin{tabular}{ c | c | c }
	Model & A & B  \\
	\hline
	$\ \Delta_V\ $  						& $\pm 1$  				& $\pm 1$	\\
	$\ V_e\ |\Delta_V|\ $ 			& $1000$  				& $1000$	\\
	$\ \mu_{\nu}\ |\Delta_V|\ $ & $10\,000$ 			& $\ 10\,000\ {\rm e}^{-r/10}$ \\
	$\ \alpha\ $    						& $\ 1.3-0.0048\,r\ $ & $4/3$	\\
	\hline
	\end{tabular}
\caption{Chosen parameter values for the Symmetric model (Model A) and the Standard model (Model B): vacuum scale, $\Delta_V$ (+1 for normal, -1 for inverted neutrino mass hierarchy), 
background matter potential, $V_e$, neutrino-neutrino interaction strength, $\mu_{\nu}$, and the 
$\nu_e$ and $\bar{\nu}_e$ asymmetry factor, $\alpha$.}
\label{tab:models}
\end{table}

In our calculations, we assume neutrinos to be produced as pure flavor states. We follow the evolution of (anti)neutrinos by solving the evolution equations in Eq.~\eqref{eq:evoeqs} with the following initial conditions for the neutrino and antineutrino density matrices in Eq.~\eqref{eq:rhos}:
\begin{equation}
\label{eq:rhoini}
	\rho^0 \equiv \rho(r=0) =
		\begin{pmatrix}
		1 & 0 \\
		0 & 0
	\end{pmatrix}
  =	\bar{\rho}(r=0) \equiv \bar{\rho}^0 \ .
\end{equation}
Taking Eq. (\ref{eq:rhoini}) and substituting it into Eq. (\ref{eq:Vnu}) provides information about what
$V_\nu$ would be in the absence of any oscillation, i.e. $V_\nu^{unosc} = \mu_\nu(1-\alpha)$.

\begin{figure}[th!]
		\includegraphics[width=0.9\textwidth]{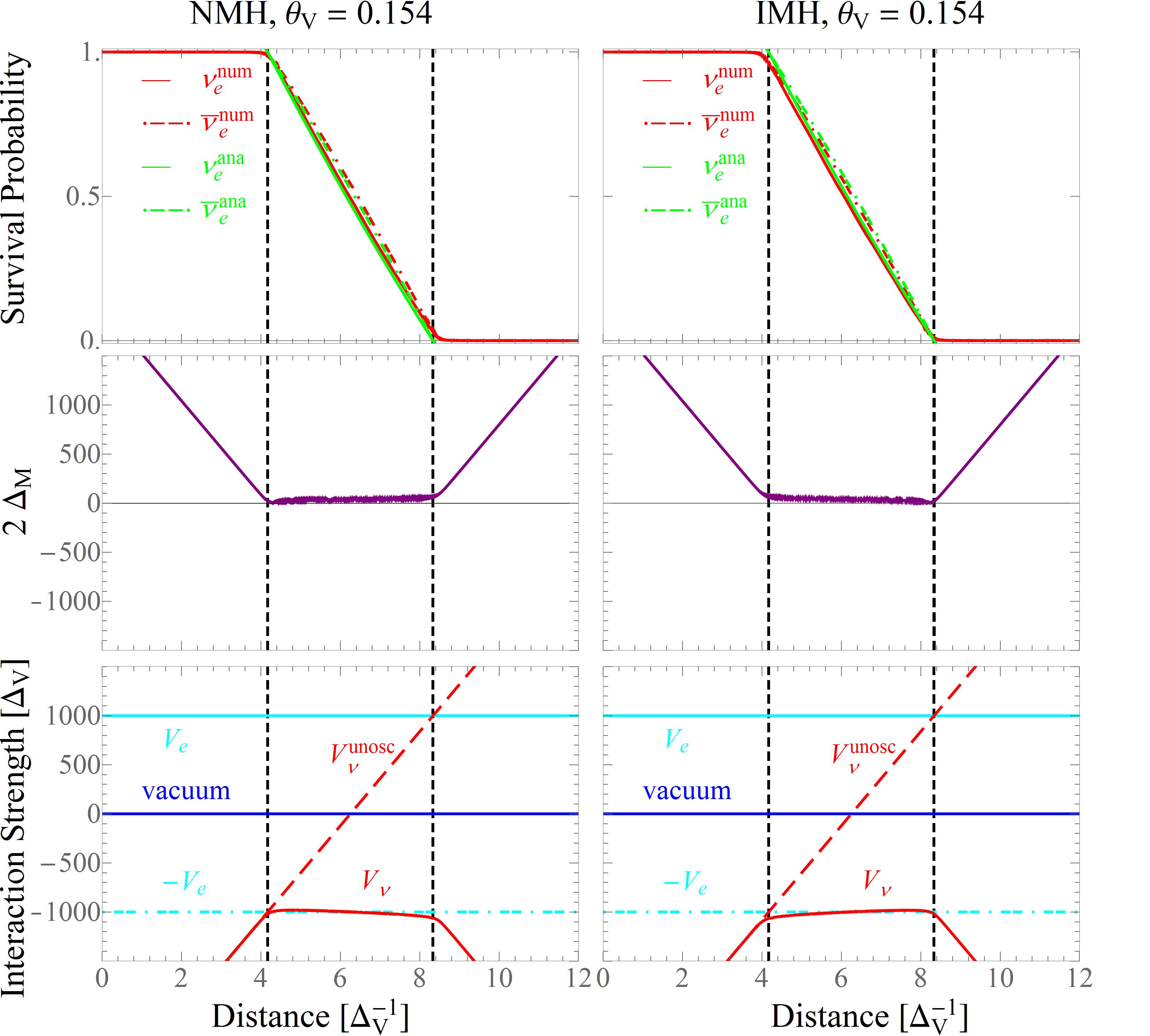}
  \caption{(Color online) Comparison of numerical results with analytical predictions in the Symmetric MNR model assuming vacuum mixing angle $\theta_V = 0.154$  (model A). 
Left (right) figures display the results in normal (inverted) mass hierarchy, $\Delta_V = +1 (-1)$. {\bf Top figures:} Survival probabilities for $\nu_e$ (solid) and $\bar{\nu}_e$ (dashed-dotted lines) with a comparison of numerical results (red) with analytical prediction given by Eq.~\eqref{eq:anal} (green lines). {\bf Middle figures:} Comparison of MNR assumption, $\Delta_M = 0$ (light gray line), with numerically computed in-medium 
eigenvalue difference (purple line), utilizing Eq.~\eqref{eq:deltaM} and numerical results for the neutrino and antineutrino densities. {\bf Bottom figures:} Contributions to the total neutrino Hamiltonian, Eq.~\eqref{eq:HF}: Vacuum (blue), background matter potential, $V_e$ (cyan), neutrino self interaction potential, $V_{\nu}$ (solid red line) as well as $V_{\nu}^{unosc}$ (dashed red line). During MNR transition, neutrino-neutrino potential actively cancels the background matter contribution and should be compared with $-V_e$. Vertical dashed lines represent the resonance locations according to Eq. (\ref{eq:MNRcond}).
	}
	\label{fig:thetaLargemodelA}
\end{figure}

\begin{figure}[th!]
\includegraphics[width=0.9\textwidth]{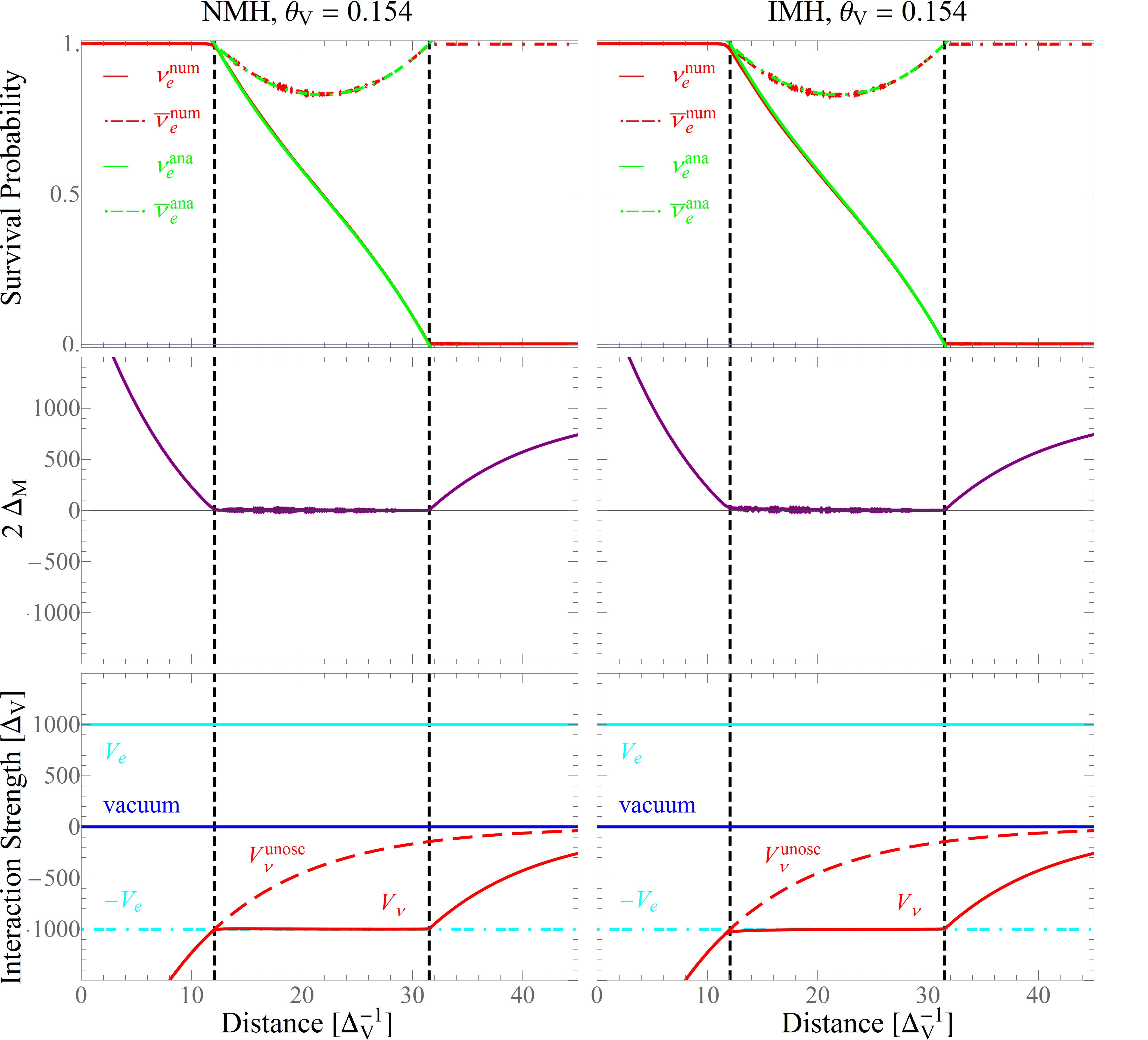}
		\caption{Comparison of numerical results with analytical predictions in the Standard MNR model (model B) assuming vacuum mixing angle $\theta_V = 0.154$. Lines are as described in Fig.~\ref{fig:thetaLargemodelA}.
	}
	
	\label{fig:standthetaLarge}
\end{figure}

Having set up both models we have one last choice to make which is the vacuum mixing angle.  In this section we take the value of the vacuum mixing angle to be consistent with $\theta_{13}$: $\theta_V = 0.154$~\cite{Agashe:2014kda}.  We can now turn to our results, which are presented in Figs.~\ref{fig:thetaLargemodelA} and \ref{fig:standthetaLarge}. In the top panels of these figures we show the comparison of analytic and numerical results for the Symmetric (Fig.~\ref{fig:thetaLargemodelA}) and Standard (Fig. \ref{fig:standthetaLarge}) cases for the inverted and normal hierarchies. In all four top panels, the analytical results match well with the numerical results for the survival probabilities and the transition starts at the resonance location as predicted.

The analytical results were derived from the assumption that the difference between the instantaneous in-medium eigenvalues was close to zero. We verify the appropriateness of this assumption in the middle panels in these figures where we plot the difference of the in-medium mass eigenvalue.  We see that, this difference, 2 $\Delta_M$, approaches zero. Another prediction, given in Eq.~\eqref{eq:diagmini}, was that the neutrino-neutrino potential $V_\nu$ will mirror the matter potential $V_e$. As can be seen from the bottom panels in these four figures, $V_\nu$ deviates strongly from the value it would take on if no oscillation occurred, $V_\nu^{unosc}$, and tracks $V_e$ as expected.

\subsection{Sensitivity to the Vacuum Mixing - Linearized (In)stability Analysis}
\label{sec:stab}

The resonance locations, Eq.~\eqref{eq:MNRcond}, for  MNR transitions have a modest dependence on the vacuum mixing angle in regions well above the vacuum scale. However, the 
adiabaticity parameter at the location of resonance, $\gamma_R$ in Eq.~\eqref{eq:gammaR}, scales as $\sin^2 2\theta_V$.  Thus the presence or absence of the resonance transition is dependent on the vacuum mixing angle.  Either reducing the mixing angle or steepening the neutrino and/or matter profiles will shut off the MNR transition.  In this section we consider the former scenario.

\begin{figure}[th!]
\includegraphics[width=0.9\textwidth]{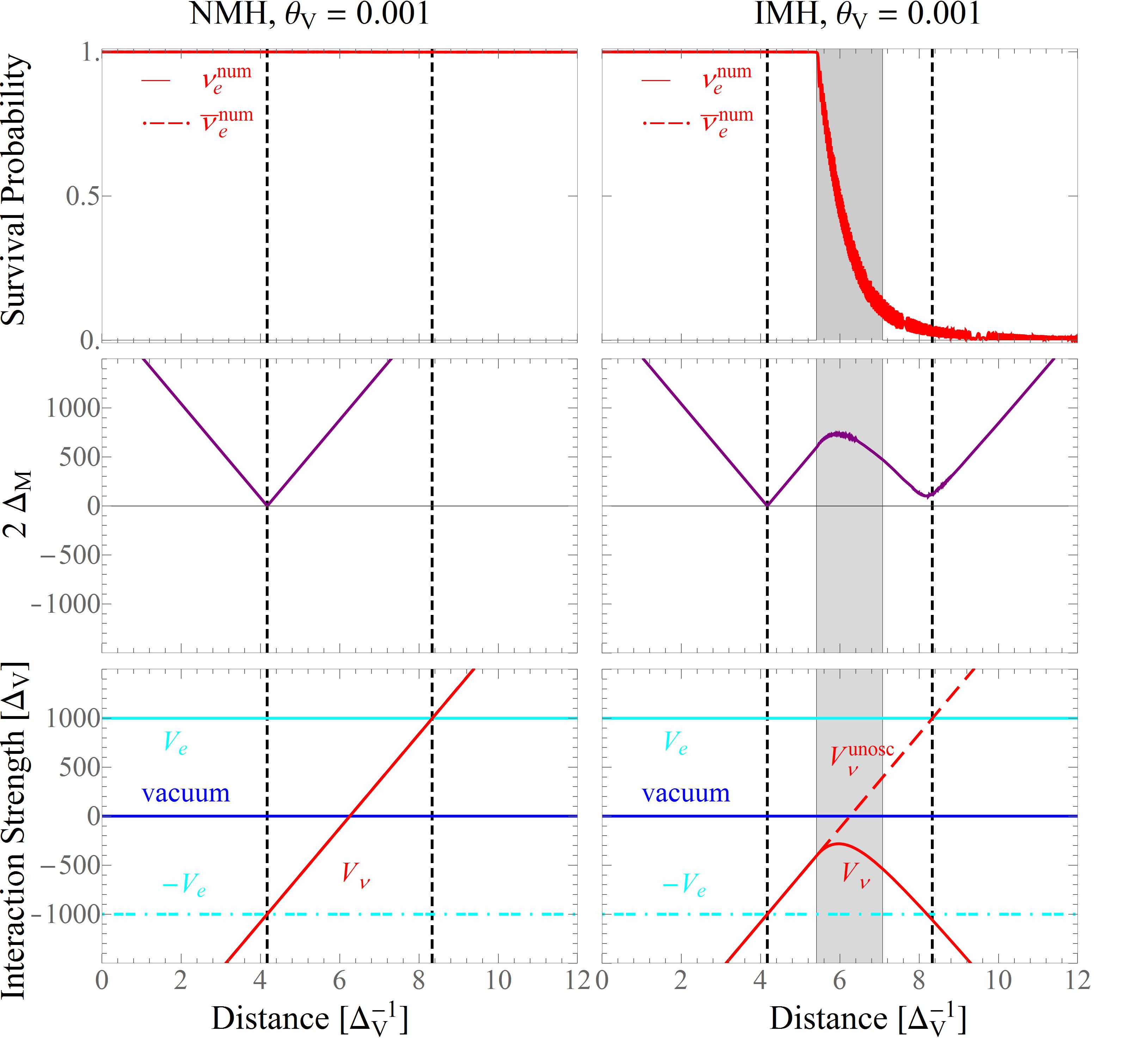}
		  \caption{(Color online) Comparison of numerical results with analytical predictions in the Symmetric model assuming small vacuum mixing angle $\theta_V = 0.001$ (other parameters as in Model A described in Table~\ref{tab:models}). The figure labels and lines plotted are as in Fig.~\ref{fig:thetaLargemodelA}. Due to the very small vacuum mixing no flavor transitions take place at the MNR resonance locations (vertical dashed lines).  The shaded area represents the instability region according to Eq.~\eqref{eq:stabcond} (visible only in inverted mass hierarchy (IMH) as the normal mass hierarchy (NMH) is flavor stable, see Section~\ref{sec:stab} for more details).  The instability analysis correctly predicts the location of the self-induced nutation type neutrino flavor transitions. 
	}
	\label{fig:thetaSmall}
\end{figure}

\begin{figure}[th!]
\includegraphics[width=0.9\textwidth]{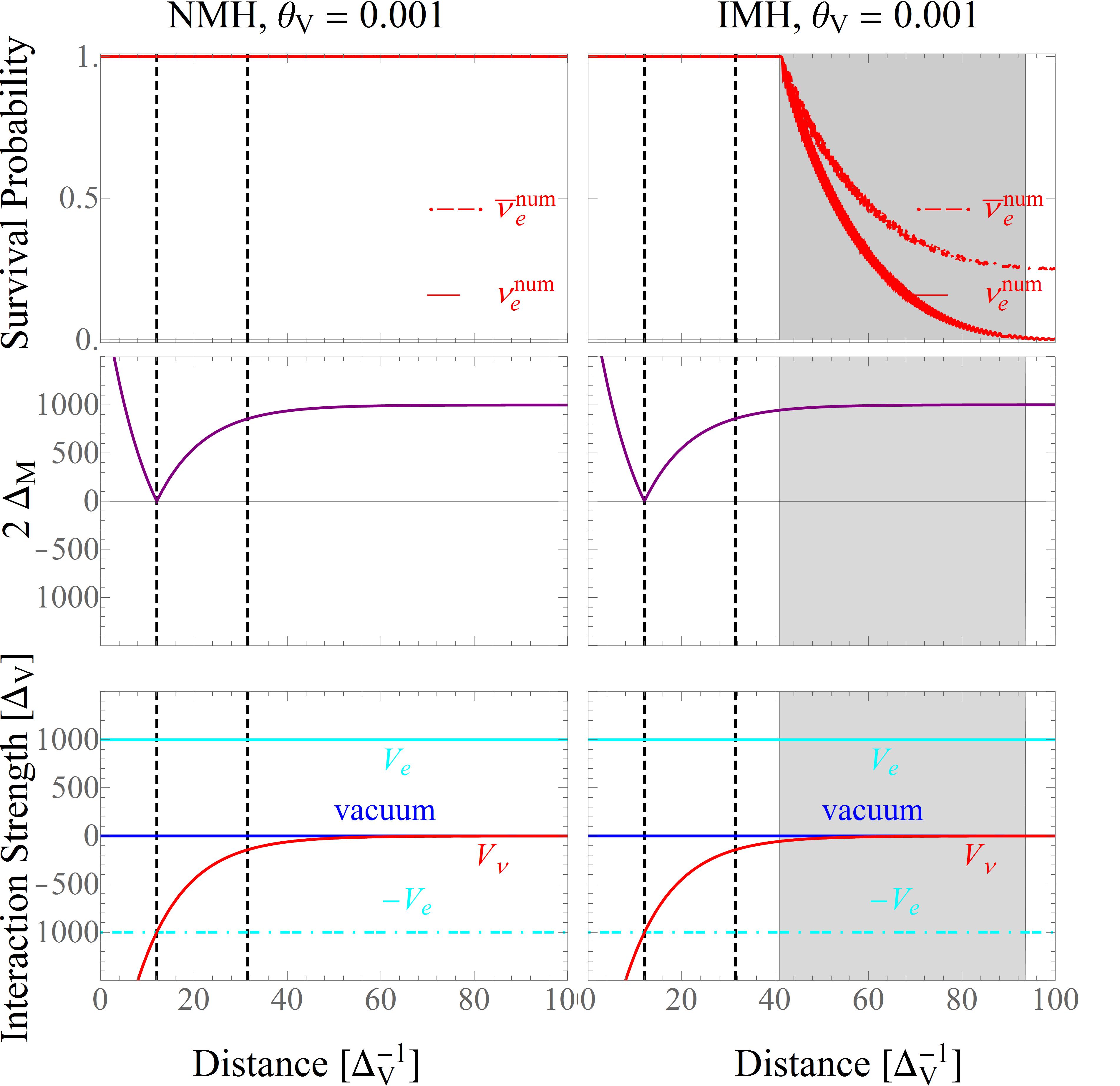}
    \caption{(Color online) Comparison of numerical results with analytical predictions in the Standard MNR model assuming small vacuum mixing angle $\theta_V = 0.001$ (other parameters as in model B described in Table~\ref{tab:models}). The figure labels and lines are as in Fig.~\ref{fig:thetaSmall}. The instability region present in IMH (the shaded area) appears after the MNR region (between vertical dashed lines). 
	}
	\label{fig:StandardthetaSmall}
\end{figure}

In Figs. \ref{fig:thetaSmall} and \ref{fig:StandardthetaSmall} we present the results for our Symmetric (Model A in Table~\ref{tab:models}) and Standard MNR system (Model B) with a reduced vacuum mixing of $\theta_V = 0.001$.  It can be seen from the top panels of these two figures that no transition begins at the initial MNR resonance position (first dashed line).  From the middle panels, we also see that the difference of the in-medium mass eigenvalues does not hover around zero between the dashed lines, as it would during a MNR transition.  Due to the small value of the vacuum mixing, the resonances are non-adiabatic.  
However, looking again at the top panels, we see that in the inverted hierarchy transitions do take place at other locations. These are self-induced nutation type transitions similar to the type found in core collapse supernovae.

The onset of the self-induced effects has been related to the presence of an instability~\cite{Sawyer:2008zs,Banerjee:2011fj,Vaananen:2013qja}. In the following, we will apply the linearized stability analysis procedure outlined in Ref.~\cite{Vaananen:2013qja} to describe conditions for neutrino flavor instability in systems which have MNR resonances. 

In order to study the stability of the systems we consider here, it is sufficient to consider the following $2\times 2$ stability matrix (see Appendix): 
\begin{equation}
\label{eq:stab22}
{\bf S} = \left(
	\begin{array}{cc}
		A_{12} & B_{12} \\
		\bar{B}_{21} & \bar{A}_{21}
	\end{array}
		\right) \quad \ ,
\end{equation}
with 
\begin{equation}
\label{eq:S2elem}
\begin{aligned}
 A_{12} =& {H}^0_{11} - {H}^0_{22} + ({\rho}_{22}^0 - {\rho}_{11}^0) 
\frac{\partial {H}_{12}}{\partial \rho_{12}} \ , \\ 
B_{12} =& ({\rho}_{22}^0 - {\rho}_{11}^0) \frac{\partial {H}_{12}}{\partial \bar{\rho}_{21}} \ , \\ 
\bar{A}_{21} =&  
\bar{H}^0_{22} - \bar{H}^0_{11} + (\bar{\rho}_{11}^0 - \bar{\rho}_{22}^0) 
\frac{\partial \bar{H}_{21}}{\partial \bar{\rho}_{21}}
\ , \\ 
\bar{B}_{21} =&
(\bar{\rho}_{11}^0 - \bar{\rho}_{22}^0) 
\frac{\partial \bar{H}_{21}}{\partial \rho_{12}}  \ , 
\end{aligned}
\end{equation}
where
$\rho^0$ is given by Eq.~\eqref{eq:rhoini} and the system is initially described by the Hamiltonian $H^0 = H_M$ with $\rho = \rho^0$, that is, by Eq.~\eqref{eq:HM} with neutrino and antineutrino density matrix elements given according to Eq.~\eqref{eq:rhoini}:
\begin{equation}
\begin{aligned}
 	\accentset{(-)}{H}^{\,0}_{22} - \accentset{(-)}{H}^{\,0}_{11} =& \ 2 \accentset{(-)}{\Delta}_M^{\,0}
	= 2\sqrt{\left(\Delta_V \cos2\theta_V\ \accentset{(+)}{-}\ (V_e + \mu_{\nu}(1-\alpha)) \right)^2 + \Delta_V^2 \sin^2 2 \theta_V }  \ , \\
	\frac{\partial {H}_{12}}{\partial \rho_{12}} =& \ \mu_{\nu} \ ,\quad \frac{\partial {H}_{12}}{\partial \bar{\rho}_{21}} = -\mu_{\nu}\alpha \ , \\
	\frac{\partial \bar{H}_{21}}{\partial \rho_{12}} =& -\mu_{\nu}  \ , 
	\frac{\partial \bar{H}_{21}}{\partial \bar{\rho}_{21}} = \ \mu_{\nu}\alpha \ .  
\end{aligned}
\end{equation}
Then, the elements of the stability matrix become (in the limit $\theta_V \rightarrow 0$):
\begin{equation}
\begin{aligned}
A_{12} &= -2 \left( \Delta_V - V_e - 
      \mu_{\nu} (1 - \alpha) \right) - \mu_{\nu} \ , \\
\bar{A}_{21} &= 2 \left( \Delta_V + V_e + 
      \mu_{\nu} (1 - \alpha) \right) + \alpha \mu_{\nu} \ ,  \\
B_{12} &= \alpha \mu_{\nu} \ , \\
\bar{B}_{21} &= -\mu_{\nu} \ .
\end{aligned}
\end{equation}
Instability conditions are obtained by solving the eigenvalues, $\lambda$, of the stability matrix: 
\begin{equation}
\label{eq:eigstab1}
 |{\bf S} - \lambda | = 0 \ .
\end{equation}
A complex value for the eigenvalue indicates an unstable mode. 
The stability matrix has eigenvalues 
\begin{equation}
\label{eq:eigstab}
 \lambda = \frac{1}{2}\left( A_{12} + \bar{A}_{21} \pm \sqrt{(A_{12} - \bar{A}_{21})^2 - 4 B_{12}\bar{B}_{21}} \right)  \ .
\end{equation}

The system is unstable if eigenvalues become imaginary, that is, if 
\begin{equation}
\label{eq:stabcond}
 (A_{12} - \bar{A}_{21})^2 < 4 B_{12}\bar{B}_{21}  \ .
\end{equation}
This stability analysis predicts the conditions under which the system becomes unstable to small perturbations. In the context of self-induced collective effects, the stability analysis gives the conditions for the on-set of self-induced flavor transformations. In the normal mass hierarchy (NMH), eigenvalues of the stability matrix in Eq.~\eqref{eq:eigstab} are always real. Hence, the system is flavor stable and no flavor transformation occur. In the inverted mass hierarchy (IMH), eigenvalues of the stability matrix become imaginary. The region where the eigenvalues become imaginary for Models A and B is shown as the shaded region in Figs. \ref{fig:thetaSmall} and \ref{fig:StandardthetaSmall}. The system has an unstable region and exhibits self-induced nutation type transformations in this region.

\section{Conclusions}
\label{sec:conc}

In this manuscript we have studied neutrino systems of interest in astrophysical environments that have the potential to exhibit Matter Neutrino Resonances (MNRs).  An example of such a system is a merging compact object.  MNR transitions may occur closer to the cores of these objects than other types of flavor transformation and so are of keen interest for nucleosynthesis and perhaps also the dynamics of these environments. 

Using models of monoenergetic neutrino gases, we have have provided general resonance conditions applicable to MNR transitions. We have shown that the MNR criteria are obtained as a consequence of the small separation of the (anti)neutrino in-medium energy eigenstates during the transition.  These criteria lead to analytical expressions for (anti)neutrino survival probabilities that accurately describe the neutrino flavor evolution during MNR transitions. Also, we have discussed how MNR transitions can be explained as an adiabatic evolution of the in-medium neutrino energy eigenstates, that is, (anti)neutrinos stay on their in-medium eigenstates throughout the transformation. 

While originally discussed as a mechanism which leaves neutrinos converted but antineutrinos in their original configuration, we have discussed how some types of MNR transitions can fully convert both neutrinos and antineutrinos.  Although the final flavor content is different, these symmetric transitions are described in the same way as standard transitions.

In the systems we have studied, the presence of a MNR transition suppressed the type of self-induced flavor transformation that has been studied in the context of core collapse supernovae.  This is because the required initial conditions at the instability point were not met.  In order for MNR transitions to take place, the vacuum mixing has to be sufficiently large. In our example models, the measured value of the `reactor' neutrino mixing angle $\theta_{13}$ is sufficient to trigger the MNR transitions.   However, smaller angles and/or steeper potentials will suppress these MNR transitions.

 If MNR resonances are ineffective, neutrinos can still undergo self-induced flavor transitions. We have applied a general linearization procedure to our Symmetric and Standard MNR models and constructed a  stability matrix allowing us to study the flavor stability in these models.  In symmetric scenarios, the instability region lies within the MNR resonance region.  Thus, if the MNR transition is suppressed, self-induced flavor transition will still occur within the resonance region in the case of the inverted hierarchy.  In standard scenarios, the instability region comes after the MNR resonance region.

In an astrophysical system, the exact location of transitions is of significance in determining the impact on dynamical evolution of the environment and conditions for nucleosynthesis. Further investigations are required in order to study the behavior of neutrinos encountering Matter Neutrino Resonances in more realistic scenarios.

\begin{acknowledgments}
We thank J. Kneller and A. Malkus for useful discussions.  This work was supported in part by U.S. DOE Grants No. DE-FG02-02ER41216, DE-SC0004786 and DE-SC0006417.  During completion of this manuscript another work investigating the underlying physics of the Standard MNR transitions appeared on the archiv \cite{Wu:2015fga}.
\end{acknowledgments}

\section{Appendix: Linearized (In)stability Analysis}

The (anti)neutrino flavor evolution of our models is described by Eq.~\eqref{eq:evoeqs}. Therefore, the evolution of an $ij$ element of the (anti)neutrino density matrix is given by
\begin{equation}
\label{eq:evoeqsij}
	\begin{aligned}
	{\rm i} \frac{{\rm d} \rho_{ij}}{{\rm d} r} = \sum_k \left(H_{ik}{\rho}_{kj} - {\rho}_{ik} H_{kj} \right) \ , \\
	{\rm i} \frac{{\rm d} \bar{\rho}_{ij}}{{\rm d} r} = \sum_k \left(\bar{H}_{ik}\bar{\rho}_{kj} - \bar{\rho}_{ik} \bar{H}_{kj} \right) \ ,
	\end{aligned}
\end{equation}
where indices $ij$ refer to the $ij$ element of the corresponding matrix.
The above set of evolution equations can be linearized by considering a time dependent small amplitude variation, $\delta \rho$, around the initial configuration, $\rho^0$, and a corresponding variation of the density dependent Hamiltonian, $\delta H$, around the initial Hamiltonian $H^0$: 
\begin{equation}
\label{eq:rhoij}
\begin{aligned}
	\rho_{ij} =& \rho^0_{ij} + \delta\rho_{ij} \ , \quad  {\rm with} \quad 
	\delta \rho_{ij} = \rho_{ij}'\,{\rm e}^{-{\rm i} \omega r} + {\rm H. c.} 
	\quad {\rm and} \\
	H_{ij} =& H^0_{ij} + \delta H_{ij} \ , \ {\rm with} \quad 
	\delta h_{ij} = H_{ij}'\,{\rm e}^{-{\rm i} \omega r} + {\rm H. c.} \ ,
\end{aligned}
\end{equation}
where $\rho'_{ij}, H'_{ij}$ are the variation amplitudes and $\omega$ describes the variation frequency. 

In the case of self-induced collective neutrino effects, the small amplitude variations around the initial configuration, $\rho_{ij}'$ in Eq.~\eqref{eq:rhoij}, are induced by the off-diagonal elements of the (anti)neutrino density matrix. 
According to our convention, the off-diagonal neutrino Hamiltonian matrix
element $H_{ij}\ (i \neq j)$ depends on the neutrino density matrix element $\rho_{ij}$ and on the antineutrino density matrix element $\bar{\rho}_{ji}$. By retaining only the contributing terms, the variation amplitude of the neutrino Hamiltonian due to the variation of the (anti)neutrino densities can be written as
\begin{equation}
\label{eq:dhij}
H_{ij}' = \sum_{k<l} \left(\frac{\partial {H}_{ij}}{\partial \rho_{kl}} {\rho}_{kl}' +
\frac{\partial {H}_{ij}}{\partial \bar{\rho}_{lk}}\bar{\rho}_{lk}' \right) \ .
\end{equation}

The initial configuration is described by the in-medium Hamiltonian that is obtained by diagonalizing the flavor basis Hamiltonian at the initial time.
Neutrino mixing is modified in medium as described by Eq.~\eqref{eq:theta2M}. Hence, with large interaction potentials, the in-medium eigenstates initially coincide with flavor states. Therefore, the initial system is described by 
\begin{equation}
\label{eq:linini}
	\left[ H^0, \rho^0 \right] = 0 \ ,
\end{equation}
and the initial configuration can be written as
\begin{equation}
\label{eq:linrhoini}
	\rho^0_{ij} = \rho_i^0\delta_{ij} \ , \quad 
	H^0_{ij} = H_i^0\delta_{ij} \ .
\end{equation}

Substituting Eqs.~\eqref{eq:rhoij},~\eqref{eq:dhij} and~\eqref{eq:linrhoini} into the evolution equations, Eq.~\eqref{eq:evoeqsij}, collecting the positive frequency modes, ${\rm e}^{-{\rm i} \omega r} $ $(i<j)$, and neglecting the higher-order corrections from $\left[ \delta h, \delta\rho \right]$, one obtains the following eigenvalue equations 
\begin{equation}
\label{eq:eigeneq}
\begin{aligned}
 \omega\rho_{ij}' =& 
\sum_{k<l} \left\{ \left[ ({H}^0_{k} - {H}^0_{l})\delta_{ik}\delta_{jl} + ({\rho}_{j}^0 - {\rho}_{i}^0) \frac{\partial {H}_{ij}}{\partial \rho_{kl}}
\right] {\rho}_{kl}' + ({\rho}_{j}^0 - {\rho}_{i}^0) \frac{\partial {H}_{ij}}{\partial \bar{\rho}_{lk}} \bar{\rho}_{lk}'  \right\} \ , 
\\
\omega\bar{\rho}_{ji}' =& 
\sum_{k<l} \left\{ \left[ (\bar{H}^0_{l} - {H}^0_{l})\delta_{il}\delta_{jk} + ({\rho}_{i}^0 - {\rho}_{j}^0) \frac{\partial \bar{H}_{ji}}{\partial \bar{\rho}_{lk}} \right] \bar{\rho}_{lk}' + (\bar{\rho}_{i}^0 - {\rho}_{j}^0) 
\frac{\partial \bar{H}_{ji}}{\partial \rho_{kl}}
 {\rho}_{kl}' \right\} \ , 
\end{aligned}
\end{equation}
with eigenvalues $\omega$ and eigenvectors $\accentset{(-)}{\rho}\ '$. There exist two sets of eigenvalue equations, one for $\omega$ and another one for its complex conjugate which can be obtained by collecting the ${\rm e}^{+{\rm i} \omega^* r} $ modes.

If the eigenvalues $\omega \in {\cal R}e$, system has a stable solution with collective oscillation modes
\begin{equation}
\label{eq:rhoijRe}
	\rho_{ij} = 2 \rho_{ij}'\cos \omega_{s} r \ .
\end{equation}
where $\omega_{s}$ represents the {\it synchronized} oscillation frequency with oscillation amplitude $2\rho_{ij}'$. On the other hand, if $\omega \in {\cal I}m$, the variations can grow exponentially, indicating that the system has become unstable and the linearized equations no longer serve as a good approximation. In the context of self-induced collective neutrino effects, instability indicates the on-set of the nutation type (or {\it bipolar}) oscillations.

The linearized eigenvalue equations, Eqs.~\eqref{eq:eigeneq}, can be written in a compact matrix form by introducing {\it Stability matrix}, ${\bf S}$:
\begin{equation}
\omega \left(
	\begin{array}{cc}
		\rho' \\
		\bar{\rho}' 
	\end{array}
		\right) = {\bf S} \left(
	\begin{array}{cc}
		\rho' \\
		\bar{\rho}' 
	\end{array}
		\right)  \ , 
\end{equation}
where 
\begin{equation}
{\bf S} = \left(
	\begin{array}{cc}
		A & B \\
		\bar{B} & \bar{A}
	\end{array}
		\right) \quad \ ,
\end{equation}
with elements
\begin{equation}
\begin{aligned}
A_{ij,kl} &= ({h}^0_{k} - {h}^0_{l})\delta_{ik}\delta_{jl} + ({\rho}_{j}^0 - {\rho}_{i}^0) \frac{\partial {h}_{ij}}{\partial \rho_{kl}}
 \ , \\
\bar{A}_{ij,kl} &= (\bar{h}^0_{l} - {h}^0_{l})\delta_{il}\delta_{jk} + ({\rho}_{i}^0 - {\rho}_{j}^0) \frac{\partial \bar{h}_{ji}}{\partial \bar{\rho}_{lk}} \ ,  \\
B_{ij,kl} &= ({\rho}_{j}^0 - {\rho}_{i}^0) \frac{\partial {h}_{ij}}{\partial \bar{\rho}_{lk}} \ , \\
\bar{B}_{ij,kl} &= (\bar{\rho}_{i}^0 - {\rho}_{j}^0) 
\frac{\partial \bar{h}_{ji}}{\partial \rho_{kl}}
 \ .
\end{aligned}
\end{equation}
$\accentset{(-)}{A},\accentset{(-)}{B}$ are $N\times N$ matrices with $N = 0.5 \, n_f \times (n_f - 1) \times n_E \times n_u \times n_{ini}$ where $n_f, n_E, n_u, n_{ini}$ are the number of
neutrino families, neutrino energies, angular modes and initial conditions, respectively, while $\rho'$ and $\bar{\rho}$ are in turn $N$-dimensional
vectors of variation amplitudes.

Instability conditions are obtained by studying the eigenvalues of the stability matrix as discussed in Section~\ref{sec:stab}. 
In case of two (anti)neutrino flavors with single energy and emission angle, the system can be decomposed into two subsystems described by a subsystem with $(1,2)$ element of the neutrino density matrix linked to $(2,1)$ element of the antineutrino density matrix and another subsystem described by the complex conjugates of the corresponding elements. The two subsystems have identical stability conditions. This consideration leads to the form of the stability matrix as shown in Eq.~\eqref{eq:stab22}.

\bibliographystyle{ieeetr}
\bibliography{refs}

\begin{thebibliography}{10}

\bibitem{Palenzuela:2015dqa}
C.~Palenzuela, S.~Liebling, D.~Neilsen, L.~Lehner, O.~Caballero, {\em et~al.},
  ``{Effects of the microphysical Equation of State in the mergers of
  magnetized Neutron Stars With Neutrino Cooling},'' 2015.

\bibitem{Dessart:2008zd}
L.~Dessart, C.~Ott, A.~Burrows, S.~Rosswog, and E.~Livne, ``{Neutrino
  signatures and the neutrino-driven wind in Binary Neutron Star Mergers},''
  {\em Astrophys.J.}, vol.~690, p.~1681, 2009.

\bibitem{Deaton:2013sla}
M.~B. Deaton, M.~D. Duez, F.~Foucart, E.~O'Connor, C.~D. Ott, {\em et~al.},
  ``{Black Hole-Neutron Star Mergers with a Hot Nuclear Equation of State:
  Outflow and Neutrino-Cooled Disk for a Low-Mass, High-Spin Case},'' {\em
  Astrophys.J.}, vol.~776, p.~47, 2013.

\bibitem{Perego:2014fma}
A.~Perego, S.~Rosswog, R.~Cabezon, O.~Korobkin, R.~Kaeppeli, {\em et~al.},
  ``{Neutrino-driven winds from neutron star merger remnants},'' 2014.

\bibitem{O'Connor:2014rva}
E.~O’Connor, ``{An Open-Source Neutrino Radiation Hydrodynamics Code for
  Core-Collapse Supernovae},'' {\em Astrophys. J. Suppl.}, vol.~219, no.~2,
  p.~24, 2015.

\bibitem{Abdikamalov:2012zi}
E.~Abdikamalov, A.~Burrows, C.~D. Ott, F.~Loffler, E.~O'Connor, J.~C. Dolence,
  and E.~Schnetter, ``{A New Monte Carlo Method for Time-Dependent Neutrino
  Radiation Transport},'' {\em Astrophys. J.}, vol.~755, p.~111, 2012.

\bibitem{Tamborra:2014hga}
I.~Tamborra, G.~Raffelt, F.~Hanke, H.-T. Janka, and B.~Mueller, ``{Neutrino
  emission characteristics and detection opportunities based on
  three-dimensional supernova simulations},'' {\em Phys. Rev.}, vol.~D90,
  no.~4, p.~045032, 2014.

\bibitem{Surman:2005kf}
R.~Surman, G.~C. McLaughlin, and W.~R. Hix, ``{Nucleosynthesis in the outflow
  from gamma-ray burst accretion disks},'' {\em Astrophys. J.}, vol.~643,
  pp.~1057--1064, 2006.

\bibitem{Roberts:2010wh}
L.~Roberts, S.~Woosley, and R.~Hoffman, ``{Integrated Nucleosynthesis in
  Neutrino Driven Winds},'' {\em Astrophys.J.}, vol.~722, pp.~954--967, 2010.

\bibitem{Duan:2010af}
H.~Duan, A.~Friedland, G.~C. McLaughlin, and R.~Surman, ``{The influence of
  collective neutrino oscillations on a supernova r-process},'' {\em J. Phys.},
  vol.~G38, p.~035201, 2011.

\bibitem{Malkus:2012ts}
A.~Malkus, J.~P. Kneller, G.~C. McLaughlin, and R.~Surman, ``{Neutrino
  oscillations above black hole accretion disks: disks with electron-flavor
  emission},'' {\em Phys. Rev.}, vol.~D86, p.~085015, 2012.

\bibitem{Sigl:1992fn}
G.~Sigl and G.~Raffelt, ``{General kinetic description of relativistic mixed
  neutrinos},'' {\em Nucl. Phys.}, vol.~B406, pp.~423--451, 1993.

\bibitem{Balantekin:2006tg}
A.~B. Balantekin and Y.~Pehlivan, ``{Neutrino-Neutrino Interactions and Flavor
  Mixing in Dense Matter},'' {\em J. Phys.}, vol.~G34, pp.~47--66, 2007.

\bibitem{Friedland:2006ke}
A.~Friedland, B.~H.~J. McKellar, and I.~Okuniewicz, ``{Construction and
  analysis of a simplified many-body neutrino model},'' {\em Phys. Rev.},
  vol.~D73, p.~093002, 2006.

\bibitem{Cardall:2007zw}
C.~Y. Cardall, ``{Liouville equations for neutrino distribution matrices},''
  {\em Phys. Rev.}, vol.~D78, p.~085017, 2008.

\bibitem{Fidler:2011yq}
C.~Fidler, M.~Herranen, K.~Kainulainen, and P.~M. Rahkila, ``{Flavoured quantum
  Boltzmann equations from cQPA},'' {\em JHEP}, vol.~02, p.~065, 2012.

\bibitem{Volpe:2013jgr}
C.~Volpe, D.~Vaananen, and C.~Espinoza, ``{Extended evolution equations for
  neutrino propagation in astrophysical and cosmological environments},'' {\em
  Phys. Rev.}, vol.~D87, no.~11, p.~113010, 2013.

\bibitem{Vaananen:2013qja}
D.~Vaananen and C.~Volpe, ``{Linearizing neutrino evolution equations including
  neutrino-antineutrino pairing correlations},'' {\em Phys. Rev.}, vol.~D88,
  p.~065003, 2013.

\bibitem{Vlasenko:2013fja}
A.~Vlasenko, G.~M. Fuller, and V.~Cirigliano, ``{Neutrino Quantum Kinetics},''
  {\em Phys. Rev.}, vol.~D89, no.~10, p.~105004, 2014.

\bibitem{Serreau:2014cfa}
J.~Serreau and C.~Volpe, ``{Neutrino-antineutrino correlations in dense
  anisotropic media},'' {\em Phys. Rev.}, vol.~D90, no.~12, p.~125040, 2014.

\bibitem{MSW}
L.~Wolfenstein, ``Neutrino oscillations in matter,'' {\em Phys. Rev. D},
  vol.~17, pp.~2369--2374, May 1978.

\bibitem{ResEnhance}
S.~P. {Mikheyev} and A.~Y. {Smirnov}, ``{Resonance enhancement of oscillations
  in matter and solar neutrino spectroscopy},'' {\em Yadernaya Fizika},
  vol.~42, pp.~1441--1448, 1985.

\bibitem{SNO}
Q.~R. Ahmad {\em et~al.}, ``{Direct evidence for neutrino flavor transformation
  from neutral-current interactions in the Sudbury Neutrino Observatory},''
  {\em Phys. Rev. Lett.}, vol.~89, p.~011301, 2002.

\bibitem{kamland_msw}
K.~Eguchi {\em et~al.}, ``{First results from KamLAND: Evidence for reactor
  anti-neutrino disappearance},'' {\em Phys.Rev.Lett.}, vol.~90, p.~021802,
  2003.

\bibitem{collective}
H.~Duan and J.~P. Kneller, ``{Neutrino flavour transformation in supernovae},''
  {\em J. Phys.}, vol.~G36, p.~113201, 2009.

\bibitem{Duan:2010bg}
H.~Duan, G.~M. Fuller, and Y.-Z. Qian, ``{Collective Neutrino Oscillations},''
  {\em Ann. Rev. Nucl. Part. Sci.}, vol.~60, pp.~569--594, 2010.

\bibitem{Duan:2010bf}
H.~Duan and A.~Friedland, ``{Self-induced suppression of collective neutrino
  oscillations in a supernova},'' {\em Phys.Rev.Lett.}, vol.~106, p.~091101,
  2011.

\bibitem{Cherry:2013mv}
J.~F. Cherry, J.~Carlson, A.~Friedland, G.~M. Fuller, and A.~Vlasenko, ``{Halo
  Modification of a Supernova Neutronization Neutrino Burst},'' {\em Phys.
  Rev.}, vol.~D87, p.~085037, 2013.

\bibitem{Vlasenko:2014bva}
A.~Vlasenko, G.~M. Fuller, and V.~Cirigliano, ``{Prospects for
  Neutrino-Antineutrino Transformation in Astrophysical Environments},'' 2014.

\bibitem{Cirigliano:2014aoa}
V.~Cirigliano, G.~M. Fuller, and A.~Vlasenko, ``{A New Spin on Neutrino Quantum
  Kinetics},'' {\em Phys. Lett.}, vol.~B747, pp.~27--35, 2015.

\bibitem{Kartavtsev:2015eva}
A.~Kartavtsev, G.~Raffelt, and H.~Vogel, ``{Neutrino propagation in media:
  Flavor-, helicity-, and pair correlations},'' {\em Phys. Rev.}, vol.~D91,
  no.~12, p.~125020, 2015.

\bibitem{Malkus:2014iqa}
A.~Malkus, A.~Friedland, and G.~C. McLaughlin, ``{Matter-Neutrino Resonance
  Above Merging Compact Objects},'' 2014.

\bibitem{Malkus:2015mda}
A.~Malkus, G.~C. McLaughlin, and R.~Surman, ``{Symmetric and Standard
  Matter-Neutrino Resonances Above Merging Compact Objects},'' 2015.

\bibitem{Hannestad:2006nj}
S.~Hannestad, G.~G. Raffelt, G.~Sigl, and Y.~Y. Wong, ``{Self-induced
  conversion in dense neutrino gases: Pendulum in flavour space},'' {\em
  Phys.Rev.}, vol.~D74, p.~105010, 2006.

\bibitem{Raffelt:2007xt}
G.~G. Raffelt and A.~Y. Smirnov, ``{Adiabaticity and spectral splits in
  collective neutrino transformations},'' {\em Phys.Rev.}, vol.~D76, p.~125008,
  2007.

\bibitem{Dasgupta:2007ws}
B.~Dasgupta and A.~Dighe, ``{Collective three-flavor oscillations of supernova
  neutrinos},'' {\em Phys. Rev.}, vol.~D77, p.~113002, 2008.

\bibitem{Galais:2011jh}
S.~Galais, J.~Kneller, and C.~Volpe, ``{The neutrino-neutrino interaction
  effects in supernovae: the point of view from the matter basis},'' {\em J.
  Phys.}, vol.~G39, p.~035201, 2012.

\bibitem{Pehlivan:2011hp}
Y.~Pehlivan, A.~B. Balantekin, T.~Kajino, and T.~Yoshida, ``{Invariants of
  Collective Neutrino Oscillations},'' {\em Phys. Rev.}, vol.~D84, p.~065008,
  2011.

\bibitem{Sawyer:2008zs}
R.~Sawyer, ``{The multi-angle instability in dense neutrino systems},'' {\em
  Phys.Rev.}, vol.~D79, p.~105003, 2009.

\bibitem{Banerjee:2011fj}
A.~Banerjee, A.~Dighe, and G.~Raffelt, ``{Linearized flavor-stability analysis
  of dense neutrino streams},'' {\em Phys.Rev.}, vol.~D84, p.~053013, 2011.

\bibitem{Sarikas:2011am}
S.~Sarikas, G.~G. Raffelt, L.~Hudepohl, and H.-T. Janka, ``{Suppression of
  Self-Induced Flavor Conversion in the Supernova Accretion Phase},'' {\em
  Phys.Rev.Lett.}, vol.~108, p.~061101, 2012.

\bibitem{Saviano:2012yh}
N.~Saviano, S.~Chakraborty, T.~Fischer, and A.~Mirizzi, ``{Stability analysis
  of collective neutrino oscillations in the supernova accretion phase with
  realistic energy and angle distributions},'' {\em Phys.Rev.}, vol.~D85,
  p.~113002, 2012.

\bibitem{Mirizzi:2011tu}
A.~Mirizzi and P.~D. Serpico, ``{Instability in the Dense Supernova Neutrino
  Gas with Flavor-Dependent Angular Distributions},'' {\em Phys.Rev.Lett.},
  vol.~108, p.~231102, 2012.

\bibitem{Mirizzi:2012wp}
A.~Mirizzi and P.~D. Serpico, ``{Flavor Stability Analysis of Dense Supernova
  Neutrinos with Flavor-Dependent Angular Distributions},'' {\em Phys.Rev.},
  vol.~D86, p.~085010, 2012.

\bibitem{Sawyer:2015dsa}
R.~F. Sawyer, ``{Neutrino cloud instabilities just above the neutrino sphere of
  a supernova},'' 2015.

\bibitem{Sarikas:2012ad}
S.~Sarikas, D.~d.~S. Seixas, and G.~Raffelt, ``{Spurious instabilities in
  multi-angle simulations of collective flavor conversion},'' {\em Phys.Rev.},
  vol.~D86, p.~125020, 2012.

\bibitem{Sarikas:2012vb}
S.~Sarikas, I.~Tamborra, G.~Raffelt, L.~Hudepohl, and H.-T. Janka, ``{Supernova
  neutrino halo and the suppression of self-induced flavor conversion},'' {\em
  Phys.Rev.}, vol.~D85, p.~113007, 2012.

\bibitem{Duan:2013kba}
H.~Duan, ``{Flavor Oscillation Modes In Dense Neutrino Media},'' {\em Phys.
  Rev.}, vol.~D88, p.~125008, 2013.

\bibitem{Raffelt:2013rqa}
G.~Raffelt, S.~Sarikas, and D.~d.~S. Seixas, ``{Azimuth-angle flavor
  instability of supernova neutrino fluxes},'' 2013.

\bibitem{Chakraborty:2014lsa}
S.~Chakraborty, G.~Raffelt, H.-T. Janka, and B.~Mueller, ``{Supernova
  deleptonization asymmetry: Impact on self-induced flavor conversion},'' 2014.

\bibitem{Dasgupta:2015iia}
B.~Dasgupta and A.~Mirizzi, ``{Temporal Instability Enables Neutrino Flavor
  Conversions Deep Inside Supernovae},'' 2015.

\bibitem{Chakraborty:2015tfa}
S.~Chakraborty, R.~S. Hansen, I.~Izaguirre, and G.~Raffelt, ``{Self-induced
  flavor conversion of supernova neutrinos on small scales},'' 2015.

\bibitem{Giunti:2006fr}
C.~Giunti, ``{Neutrino Flavor States and Oscillations},'' {\em J. Phys.},
  vol.~G34, pp.~R93--R109, 2007.

\bibitem{Duan:2005cp}
H.~Duan, G.~M. Fuller, and Y.-Z. Qian, ``{Collective neutrino flavor
  transformation in supernovae},'' {\em Phys. Rev.}, vol.~D74, p.~123004, 2006.

\bibitem{Agashe:2014kda}
K.~A. Olive {\em et~al.}, ``{Review of Particle Physics},'' {\em Chin. Phys.},
  vol.~C38, p.~090001, 2014.

\bibitem{Wu:2015fga}
M.-R. Wu, H.~Duan, and Y.-Z. Qian, ``{Physics of neutrino flavor transformation
  through matter-neutrino resonances},'' 2015.

\end{thebibliography}

\end{document}